\documentclass[12pt]{article}
\usepackage[dvips]{graphicx}

\title{Regularization for zeta functions I}
\author{}

\begin{document}
{\pagestyle{empty}
\rightline{September 2006}
\rightline{~~~~~~~~~}
\vskip 1cm
\centerline{\Large \bf Regularization for zeta functions with physical applications I}
\vskip 1cm
\centerline{{Minoru Fujimoto${}^{a)}$
}and
            {Kunihiko Uehara${}^{b)}$
}}
\vskip 1cm
\centerline{${}^{a)}${\it Seika Science Research Laboratory,
Seika-cho, Kyoto 619-0237, Japan}}
\centerline{${}^{b)}${\it Department of Physics, Tezukayama University,
Nara 631-8501, Japan}}
\vskip 5mm
(Received
\vskip 2cm
\centerline{\bf Abstract}
\vskip 0.2in
  We propose a regularization technique and apply it to the Euler product of 
zeta functions, mainly of the Riemann zeta function, to make unknown some clear.
In this paper that is the first part of the trilogy, we try to demonstrate 
the Riemann hypotheses by this regularization technique and show conditions 
to realize them. 
In part two, we will focus on zeros of the Riemann zeta function 
and the nature of prime numbers  
in order to prepare ourselves for physical applications in the third part.

\vskip 0.4cm\noindent
PACS number(s): 02.30.-f, 02.30.Gp, 05.40.-a

\hfil
\vfill
\newpage}
\setcounter{equation}{0}
\addtocounter{section}{0}
\section*{I. INTRODUCTION}
\hspace{\parindent}
Regularizations by way of the zeta function have been successful with 
some physical applications so far, such as the infrared divergence in QED and 
the Casimir effect\cite{Leeuwen}, and also 
the random matrix theory\cite{Sarnak} is discussed 
associated with the zeta function. 
It is also well known that the Riemann hypothesis associated with 
the Riemann zeta function has been remained to be proved. 
We propose a regularization technique and apply this regularization 
to the Euler product of zeta functions, 
which seems to be essential to clarify the Riemann hypotheses, mainly of 
the Riemann zeta function. 
And we try to make the difficulties of the Riemann hypothesis clear by 
this regularization technique within the elementary mathematics and 
we show some conditions to perform demonstrations of the Riemann hypotheses. 

The definition of the Riemann zeta function is 
\begin{equation}
  \zeta(z)=\sum_{n=1}^\infty\frac{1}{n^z}
          =\prod_{k=1}^\infty\left(1-\frac{1}{{p_k}^z}\right)^{-1}
\label{e101}
\end{equation}
for $\Re z>1$, where the right side is the Euler product representation and 
$p_k$ is the $k$-th prime number. 
Surely the expression for $\Re z>0$ such as
\begin{equation}
  \zeta(z)=\frac{1}{1-2^{1-z}}\sum_{n=1}^\infty\frac{(-1)^{n-1}}{n^z}
\label{e102}
\end{equation}
is well known but an Euler product for it is not known.
We will show that the Euler product representation plays an essential role in the proof of 
the Riemann hypothesis, but there is no known Euler product representation for 
the Riemann zeta function for $0<\Re z<1$.
Thus we will regularize the Euler product representation, namely, the right side of 
Eq.(\ref{e101}), whose analytical continuation to the region
$\Re z<1$ has not been carried out. 

For this regularization, we consider a technique for regularizing 
the divergent series in section 2. And we show this regularization method 
is useful for some examples especially for asymptotic expansions. 
In \S 3 we apply this regularization method to the Riemann zeta function 
and try to demonstrate the Riemann hypothesis. 
And we deal with the analytic continuation for the Riemann zeta function 
in section 4. 
In \S 5 we discuss the prime number formula and how to get the large prime number.
The final section is devoted to concluding remarks.

\section*{II. METHOD OF THE DIPOLE CANCELLATION LIMIT}
\hspace{\parindent}
The function $f(z)$ is defined by
\begin{equation}
  f(z)=\lim_{n\rightarrow\infty}f_n(z)
\label{e201}
\end{equation}
where
\begin{equation}
  f_n(z)=\sum_{k=1}^na_kz^{k-1}=\sum_{k=1}^nb_k(z)
\label{e202}
\end{equation}
By the Cauchy-Hadamard theorem, the convergence radius $\rho$ is given by
\begin{equation}
  \frac{1}{\rho}=\overline{\lim_{n\rightarrow\infty}}|a_n|^{\frac{1}{n}}
\label{e203}
\end{equation}
Here we think about a technique to get significant $f(z)$ 
using $f_n(z)$ even when $|z|>\rho$.

For example, when $a_n=1$ for $|z|<1$, $f(z)$ converges and is given by 
\begin{equation}
  f(z)=\sum_{k=1}^\infty z^{k-1}=\frac{1}{1-z}\qquad(|z|<1)
\label{e204}
\end{equation}
For $|z|>1$ this equality is not valid, we will construct a method, namely, 
the analytic continuation for $|z|>1$.
Generally we think about function sequence $\alpha_k(z)$, 
which satisfies the equation
\begin{equation}
  (1-\alpha_k(z))b_k(z)+\alpha_{k+1}(z)b_{k+1}(z)=0.
\label{e205}
\end{equation}
This equation means that one of the internal division by 
$\alpha_k(z):(1-\alpha_k(z))$ for value of $k$-th term $b_k(z)$ 
cancels another of the internal division for $b_{k+1}(z)$ each other. 

In this case, the fact that $a_k=1$ and $b_k(z)=z^{k-1}$ reads, 
\begin{equation}
  (1-\alpha_k(z))z^{k-1}+\alpha_{k+1}(z)z^k=0.
\label{e206}
\end{equation}
Admitting Eq.(\ref{e206}) even in the case of $|z|>1$, we get
\begin{equation}
  1-\alpha_k(z)+\alpha_{k+1}(z)z=0.
\label{e207}
\end{equation}
Multiplying this equation by $z^{k-1}$ from $k=1$ to $k=n$, we get
\begin{equation}
  \alpha_{n+1}(z)z^n-\alpha_1(z)=\frac{z^n-1}{1-z}.
\label{e208}
\end{equation}
Because Eq.(\ref{e207}) is independent of $k$, 
we assume the common limit $\alpha(z)=\alpha_k(z)=\alpha_{k+1}(z)$ and put it into Eq.(\ref{e208}),
\begin{equation}
  1-\alpha(z)+\alpha(z)z=0
\label{e209}
\end{equation}
\begin{equation}
  \alpha(z)=\frac{1}{1-z}=\alpha_k(z)=\alpha_{k+1}(z).
\label{e210}
\end{equation}
Thus,
\begin{eqnarray}
  f_n(z)
  &=&\sum_{k=1}^nz^{k-1}=\frac{1-z^n}{1-z}\nonumber\\
  &=&\sum_{k=1}^nb_k(z)\nonumber\\
  &=&\sum_{k=1}^n\{\alpha_k(z)b_k(z)+(1-\alpha_k(z))b_k(z)\}\\
  &=&\alpha_1(z)b_1(z)+(1-\alpha_n(z))b_n(z)\nonumber\\
  &=&\frac{1}{1-z}+\frac{-z^n}{1-z}\nonumber
\end{eqnarray}
Here the initial term $\alpha_1(z)$ will be finite in the limit of 
$n\rightarrow\infty$.
For the general case, we will use the solution $\alpha_k(z)$ to the function equation (\ref{e205})
to get the finite part of the limit value $\alpha_1(z)b_1(z)$.
We call this technique the method of the dipole cancellation limit, 
and Eq.(\ref{e205}) dipole equation. 
The dipole equation is the equation to determine the form 
of function $\alpha_k(z)$.

From the fact above, asymptotic expansions or partial integrals 
of divergent sequences give of 
$$
  \alpha_k(z)=\alpha_{k+1}(z)=\alpha_{k+1}
$$
\begin{equation}
  \tilde{f}~_n(z)=\sum_{k=1}^{n-1}b_k(z)+\alpha_nb_n(z)
\end{equation}
in the case that $\displaystyle{\lim_{n\rightarrow\infty}\tilde{f}_n(z)}$ 
converge, this convergence value will be the finite value of $f(z)$. 
The condition 
$\displaystyle{\lim_{k\rightarrow\infty}\alpha_k(z)=\alpha(z)\neq0}$ 
means $b_k(z)$ is like equal ratio series, and in this case the order of 
the dipole function becomes $O(\alpha(z)b_1(z))$.

When $\alpha_k(z)\neq0$, we can apply the discussion above. 
For example, $\displaystyle{f(z)=\lim_{n\rightarrow\infty}f_n(z)}$ in which
\begin{equation}
  f_n(z)=\sum_{k=1}^\infty\frac{(k-1)!}{z^{k-1}}
\end{equation}
Eq.(\ref{e205}) gives,
\begin{equation}
  (1-\alpha_k(z))\frac{(k-1)!}{z^{k-1}}
  +\alpha_{k+1}(z)\frac{k!}{z^k}=0
\end{equation}
Actually in this case
it is difficult to get the exact solution to the dipole equation
\begin{equation}
z-z\alpha_k(z)+k\alpha_{k+1}(z)=0,
\end{equation}
but the solution will be
\begin{eqnarray}
f(z)&=&\alpha_1(z)\nonumber\\
  &=&\frac{z}{e^z}\left(\log z+\sum_{k=1}^\infty\frac{z^k}{k!k}\right)\nonumber\\
  &=&\frac{z}{e^z}\left(\log z+\sum_{k=1}^\infty\frac{z^k}{(k+1)!-k!}\right)\nonumber\\
  &=&\frac{1}{e^z}\left(z\log z+\sum_{k=1}^\infty
    \frac{z^{k+1}}{(k+1)!(1-\frac{1}{k+1})}\right)\\
  &\le&\frac{1}{e^z}\left(z\log z+\sum_{k=1}^\infty
    \frac{2z^{k+1}}{(k+1)!}\right)\nonumber\\
  &=&\frac{1}{e^z}\{z\log z+2(e^z-1-z)\}\nonumber\\
  &=&2+\frac{z\log z-2-2z}{e^z},\nonumber
\end{eqnarray}
which reads $\displaystyle{\lim_{n\rightarrow\infty}f_n(z)}<2$, namely, finite.

For an example of the case with the exception of equal ratio series, 
we show the harmonic series
\begin{equation}
  f_n(z)=\sum_{k=1}^n\frac{1}{k}.
\label{e227}
\end{equation}
From Eq.(\ref{e205}), we get the dipole equation
\begin{equation}
  \frac{\alpha_k(z)}{k}-\frac{\alpha_{k+1}(z)}{k+1}=\frac{1}{k},
\label{e228}
\end{equation}
then we get the solution $\alpha_k(z)=-k\psi(k)$,
where $\displaystyle{\psi(z)=\frac{d}{dz}\log \Gamma(z)=\frac{\Gamma'(z)}{\Gamma(z)}}$ and is called
digamma function.
Therefore $\alpha_k(z)=-k\psi(k)$ is the solution to Eq.(\ref{e228}), 
the dipole function is 
\begin{equation}
  \alpha_1(z)=-1\cdot\psi(1)=-(-\gamma)=\gamma,
\end{equation}
which is reduced to the constant.
It gives the dipole value in the case such that the dipole equation 
can be solved, but it is difficult to give the exact solution 
to the dipole equation generally. 
For another example of this regularization technique, we show the application 
of the this method to the Riemann zeta function of the summation type, 
the middle side of Eq.(\ref{e101}), in Appendix. 
The similar kind of regularization can be formulated by the continued fraction, 
but here we do not go into this way.

\section*{III. PARAMETRIZATION FOR ZETA FUNCTION}
\hspace{\parindent}
We parametrize a complex variable $z$ by two real variable as 
$\displaystyle{z=s(\frac{1}{2}+it)}$, so the Euler product representation 
of the Riemann zeta function is expressed by
\begin{equation}
  \zeta(s(\frac{1}{2}+it))=\prod_{k=1}^\infty
  \left(1-\frac{1}{p_k^{s(\frac{1}{2}+it)}}\right)^{-1}
\label{e301}
\end{equation}
Symmetry properties for the complex conjugate, which is denoted by the overline,
\begin{equation}
\zeta(s(\frac{1}{2}-it))=\zeta(\overline{s(\frac{1}{2}+it}))
=\overline{\zeta(s(\frac{1}{2}+it))}
\label{e302}
\end{equation}
will be got straightforward from the definition Eq.(\ref{e101}).

Let us think about the product $\zeta(z)\overline{\zeta(z)}$, namely, 
the square of absolute value of the zeta function, and we call hereafter this 
product the standard form of the degree two zeta function.
Euler product representation for the standard form is
\begin{eqnarray}
  \zeta(s(\frac{1}{2}+it))\overline{\zeta(s(\frac{1}{2}+it))}
  &=&\zeta(s(\frac{1}{2}+it))\zeta(s(\frac{1}{2}-it))\nonumber\\
  &=&\prod_{k=1}^\infty\left(1-\frac{1}{p_k^{s(\frac{1}{2}+it)}}\right)^{-1}
  \prod_{k=1}^\infty
            \left(1-\frac{1}{p_k^{s(\frac{1}{2}-it)}}\right)^{-1}\nonumber\\
  &=&\prod_{k=1}^\infty\left(1-\frac{2}{p_k^{s/2}}\cos(st\log p_k)
                             +\frac{1}{{p_k}^s}\right)^{-1}\\
\label{e304}
&\equiv&f(s,t)^{-1},\nonumber
\end{eqnarray}
that is, $f(s,t)=\displaystyle{\lim_{n\rightarrow\infty}f_n(s,t)}$, where
\begin{eqnarray}
  f_n(s,t)=\prod_{k=1}^n\left(1-\frac{2}{p_k^{s/2}}\cos(st\log p_k)
+\frac{1}{{p_k}^s}\right).
\label{e305}
\end{eqnarray}
The Euler product representation above is valid for $s\ge 2$, 
and we restrict our interest for $t>0$.
Graphs for the standard form calculated by the Euler product are plotted. 
The relation between local maxima of the standard form and the zeros 
of the Riemann zeta function is easily taken in for the large t. 
These figures are understood from the Hadamard product, which shows zeros 
of the Riemann zeta function except smallest one. 

\begin{figure}[t]
  \begin{center}
    \includegraphics[width=14cm,height=7cm,clip]{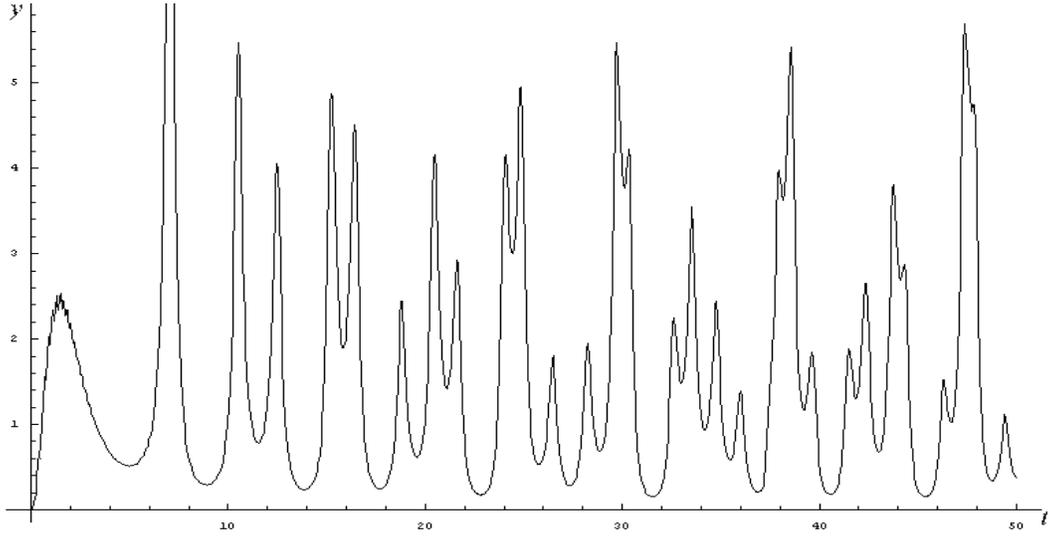}
  \end{center}
  \caption{\footnotesize The graph of $y=f_n(2,t)\;{\rm for\;}n=10^6$. 
    By the Hadamard product under the Riemann hypothesis, 
    approximate values to give local maxima are 
    $t\simeq\frac{1}{s}\sqrt{\lambda^2+\frac{1}{4}-\frac{s}{2}(1-\frac{s}{2})}
    =\frac{1}{2}\sqrt{\lambda^2+\frac{1}{4}}\;$ for $s=2$, 
    where $\zeta(1/2+i\lambda)=0$.} 
\end{figure}

\begin{figure}[!b]
  \begin{center}
    \includegraphics[width=14cm,height=7cm,clip]{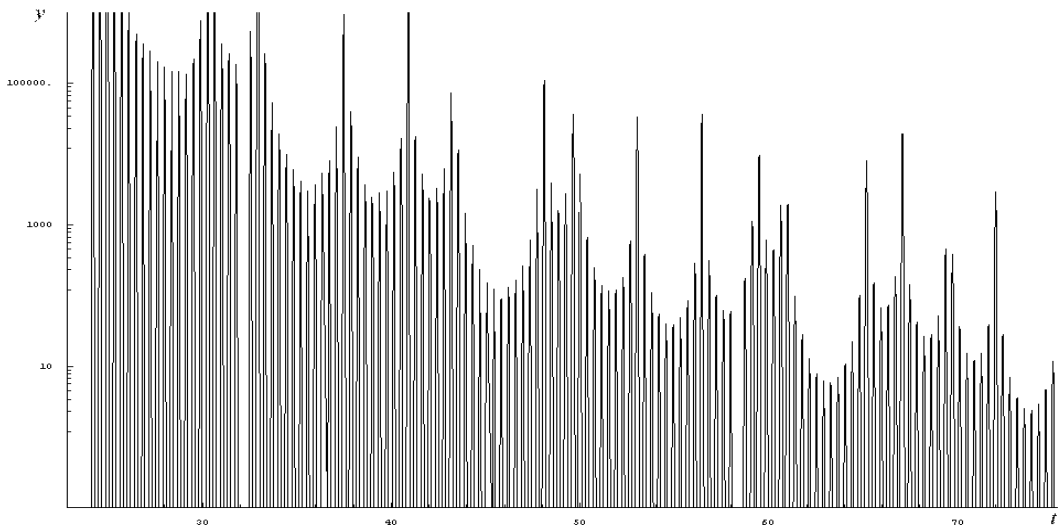}
  \end{center}
  \caption{\footnotesize The graph of $y=g_n(1,t)\;{\rm for\;}n=10^6$. The y axis 
is plotted in logarithmic scale. } 
\end{figure}

For the case of $\Re z=\frac{1}{2}$,
\begin{equation}
  g_n(s,t)\equiv\frac{f_n(s,t)}
    {\displaystyle{\prod_{k=1}^n\left(1+\frac{1}{{p_k}^s}\right)}}
\label{e306}
\end{equation}
\begin{equation}
  \lim_{n\rightarrow\infty}g_n(s,t)
  =\lim_{n\rightarrow\infty}\frac{\zeta_n(2s)}{\zeta_n(s)}
   \frac{1}{\zeta(s(\frac{1}{2}+it))\zeta(s(\frac{1}{2}-it))}
\label{e307}
\end{equation}

From these graphs above, we can see an availability of the Euler product 
even in the region of $\Re z<1$.
From the fact that "the geometric mean" is less than or is equal to 
"the arithmetic mean", we get the inequality
\begin{eqnarray}
  f(s,t)&=&
  \prod_{k=1}^\infty\left(1-\frac{2}{p_k^{s/2}}\cos(st\log p_k)
  +\frac{1}{{p_k}^s}\right)\nonumber\\
  &<&\lim_{n\rightarrow\infty}\left\{1+\frac{1}{n}
  \left(-2\sum_{k=1}^n\frac{1}{p_k^{s/2}}\cos(st\log p_k)
  +\sum_{k=1}^n\frac{1}{{p_k}^s}\right)\right\}^n
\label{e308}
\end{eqnarray}

Here we put 
\begin{equation}
  \begin{array}{rclrcl}
    B_n(s,t)&=&\displaystyle{\sum_{k=1}^n\frac{1}{p_k^{s/2}}\cos(st\log p_k)}, &
    B(s,t)&=&\displaystyle{\lim_{n\rightarrow\infty}B_n(s,t)},\\
    C_n(s)&=&\displaystyle{\sum_{k=1}^n\frac{1}{{p_k}^s}}, &
    C(s)&=&\displaystyle{\lim_{n\rightarrow\infty}C_n(s)},
  \end{array}
\label{e309}
\end{equation}
then
\begin{equation}
  f(s,t)<\lim_{n\rightarrow\infty}\left(1+\frac{-2B_n(s,t)+C_n(s)}{n}\right)^n
        =e^{-2B(s,t)+C(s)},
\label{e310}
\end{equation}
where the right side becomes finite when $B(s,t)$ and $C(s)$ are finite for $st\neq0$.
In the above deformation, we use the relation 
$\displaystyle{\lim_{n\rightarrow\infty}\left(1+\frac{A_n(x_i)}{n}\right)^n
=\lim_{n\rightarrow\infty}e^{A_n(x_i)}=e^{A(x_i)}}$, 
whose proof is shown in Appendix.

An evaluation for $C(s)$ is straightforward depending on $s$, 
\begin{equation}
C(s)=\left\{
  \begin{array}{ll}
  \displaystyle{\sum_{k=1}^\infty\frac{1}{p_k}}=+\infty & {\rm for}\;s=1 \\
  \displaystyle{\sum_{k=1}^\infty\frac{1}{{p_k}^s}<\sum_{n=1}^\infty\frac{1}{n^s}}=O(1) & 
  {\rm for}\;s>1.
  \end{array}
\right.
\label{e311}
\end{equation}

For an evaluation of $B(s,t)$, we need some deformations 
using fundamental relations\cite{Massias}
\begin{equation}
\pi(x)={\rm Li}(x)+O(xe^{c_\pi(\log x)^{1/10}}),\;\;\;c_\pi\mbox{:constant}
\label{e312}
\end{equation}
\begin{equation}
p_n=n(\log n+\log\log n-1)+o(\frac{n\log\log n}{\log n}),
\label{e313}
\end{equation}

To evaluate $B(s,t)$, we first deal with 
$B_n(s,t)=\displaystyle{\sum_{k=1}^n\frac{\cos(st\log p_k)}{p_k^{s/2}}}$. 

We put $u=p_n$, then we get
\begin{equation}
  n=\pi(u)=\int_2^u\frac{dt}{\log t}+o(\frac{u}{\log^2u}).
\label{e315}
\end{equation}
Taking\quad$\displaystyle{\frac{dn}{du}=\frac{1}{\log u}}$\quad into account,
\begin{equation}
  B_n(s,t)=\sum_{p_n}b_u(s,t)\frac{1}{\log u},
\label{e316}
\end{equation}
where $\displaystyle{b_u(s,t)=\frac{\cos(st\log u)}{u^{s/2}}}$.

We put $\log u=v$, namely, $u=e^v$ and we get using 
$\displaystyle{\frac{du}{dv}=e^v}$ 
\begin{equation}
  b_u(s,t)=b_v(s,t)\frac{du}{dv}
  =\frac{\cos(stv)}{e^{sv/2}}e^v,
\label{e317}
\end{equation}
\begin{eqnarray}
  \sum_{p_n}b_u(s,t)\frac{1}{\log u}
  &=&\sum_{\log p_n}\frac{\cos(stv)}{e^{sv/2}}e^v\frac{1}{\log u}\nonumber\\
  &=&\sum_{\log p_n}\frac{\cos(stv)}{e^{v(\frac{s}{2}-1)}}\frac{1}{v}.
\label{e318}
\end{eqnarray}
Again we put $r=stv$, namely, $v=\displaystyle{\frac{r}{st}}\;\;\;(st\neq0)$, 
$\displaystyle{\frac{dv}{dr}=\frac{1}{st}}$, then we get
\begin{eqnarray}
  \sum_{\log p_n}\frac{\cos(stv)}{e^{v(\frac{s}{2}-1)}}\frac{1}{v}
  &=&\sum_{st\log p_n}\frac{\cos r}{\displaystyle{e^{\frac{r}{st}(\frac{s}{2}-1)}
     \,\frac{r}{st}}}\frac{1}{st}+o(\frac{e^r}{v^2})\nonumber\\
  &=&\sum_{st\log p_n}\frac{\cos r}{e^{\frac{r}{st}(\frac{s}{2}-1)}\,r}
     +o(\frac{e^r}{v^2})\\
  &=&\sum_{st\log p_n}\frac{\cos r}{e^{\frac{r}{t}(\frac{1}{2}-\frac{1}{s})}\,r}
     +o(\frac{e^r}{v^2}).\nonumber
\label{e319}
\end{eqnarray}

Therefore sequences $B_n(s,t)$ which we deal with, is the summation of 
\begin{equation}
\frac{e^{ar}\cos r}{r},
\end{equation}
where $r=st\log p_n$ and we put $\displaystyle{a=(\frac{1}{s}-\frac{1}{2})\frac{1}{t}}$ 
which is positive for $1<s<2$. 

Hereafter we will invert the summation to the integral. First we use Eq.(\ref{e313})

\begin{eqnarray}
\log p_n
  &=&\log\{n(\log n+\log\log n-1)\}
   +o\left(\log(\frac{n\log\log n}{\log n})\right)\nonumber\\
  &=&\log n+o(\log n),
\end{eqnarray}

\begin{equation}
  \frac{d\log p_n}{dn}
  =\frac{1}{p_n}\frac{dp_n}{dn}
  =\frac{\log n+\log\log n}{n\log n}=\frac{1}{n}+o(\frac{1}{n}).
\end{equation}
For large $n$ with the finite value of $s$ and $t$, $\cos r$ has intervals 
in which monotonically increasing or decreasing alternately, 
therefore $B_n(s,t)$ is constrained from upper and lower limit.
Then we can always take the integral between the upper and the lower value.
The difference from the integral in the interval form $2\pi n$ to $2\pi(n+1)$
 is at most $\displaystyle{\frac{4st}{N}}$ for the large number $N$, 
since $|\cos r|\leq 1$.
Thus the difference between $B_n(s,t)$ and 
\begin{equation}
  D_n=\int_N^{st\log p_n}\frac{e^{ar}\cos r}{r}dr
\end{equation}
is finite and is evaluated by.
\begin{equation}
(1-\frac{4st}{N})D_n+O(1)<B_n(s,t)<(1+\frac{4st}{N})D_n+O(1)
\end{equation}
Therefore $D_n$ and $B_n(s,t)$ will converge or diverge simultaneously 
for the limit of $n\rightarrow\infty$.

Here we have a positive $a$ 
\begin{equation}
\lim_{n\rightarrow\infty}e^{ar}=+\infty,
\end{equation}
so $\displaystyle{\frac{e^{ar}\cos r}{r}}$ diverge positive or negative oscillatory.

For $0<\theta<2\pi$ and $m\ge 1$
\begin{equation}
 1\geq\frac{1}{1+\frac{\theta}{m}}\geq\frac{m}{m+1}\geq\frac{1}{2},
\end{equation}
\begin{equation}
  \frac{e^{am}}{m}
  \geq\frac{e^{am}}{m}\frac{1}{1+\frac{\theta}{m}}
  \geq\frac{e^{am}}{2m}.
\end{equation}
When we put 
\begin{equation}
  E=\sum_{m=1}^\infty(-1)^m\frac{e^{am}}{m}
   =\sum_{k=1}^\infty\left\{(-1)\frac{e^{a(2k-1)}}{2k-1}
   +\frac{e^{2ak}}{2k}\right\},
\end{equation}
then we can show $E>0$ for $k>\{2(1-e^{-a})\}^{-1}$. Using this
\begin{equation}
  E+O(1)
  \geq\int_N^\infty\frac{e^{ar}\cos r}{r}dr
  \geq\frac{E}{2}+O(1)
\end{equation}
\begin{equation}
  E+O(1)\geq\lim_{n\rightarrow\infty}D_n\geq\frac{E}{2}+O(1)
\end{equation}
Therefore the evaluation of the regularized $B(s,t)$ is given by 
\begin{equation}
  (1-\frac{4}{N})\frac{E}{2}+O(1)\le \lim_{n\rightarrow\infty}B_n(s,t)=B(s,t)
  \le (1+\frac{4}{N})E+O(1)
\end{equation}
and we use the relation
\begin{equation}
  \log(1+x)=-\sum_{m=1}^\infty\frac{(-1)^mx^m}{m}\;\;\;(|x|<1),
\end{equation}
we get
\begin{equation}
  -\frac{1}{2}(1-\frac{4}{N})\log(1+e^a)+O(1)\le B(s,t)\le 
    -(1+\frac{4}{N})\log(1+e^a)+O(1).
\end{equation}
After all $B(s,t)$ will be finite when the analytic continuation is performed,
\begin{equation}
  B(s,t)=O\left(-\log(1+e^{\frac{1}{t}(\frac{1}{s}-\frac{1}{2})})\right).
\end{equation}

Moreover, from the fact that "the harmonic mean" is less than or 
is equal to "the geometric mean", we can also get the behavior of the pole at $s=1$.
Here we do not go farther because the regularized Riemann zeta function is finite 
for $1<s<2$, namely, the standard form has the lower limit.

\section*{IV. ANALYTIC CONTINUATION}
\hspace{\parindent}
By using the method of the dipole cancellation limit, 
we will show the analytic continuation of 
$$B(s,t)=\lim_{n\rightarrow\infty}B_n(s,t)\quad\mbox{for }s\ge 2,\;t>0,$$ 
where
$$B_n(s,t)=\sum_{k=1}^n\frac{\cos(st\log p_k)}{{p_k}^{s/2}}.$$ 
As mentioned above section, $B_n(s,t)$ will be divergent like as the integration of 
$$\frac{e^{ar}\cos r}{r},$$
where 
$r=st\log p_n$, $a=\displaystyle{\frac{r}{t}(\frac{1}{2}-\frac{1}{s})}>0$. We will regularize
this function by using the method of the dipole cancellation limit.

\begin{eqnarray}
  b_k(r)&=&\int_{(N+k)\pi}^{(N+k+1)\pi}\frac{e^{ar}\cos r}{r}dr\\
  c_n&=&\sum_{k=1}^n\frac{\cos(st\log p_k)}{p^{s/2}}\\
  f_n(r)&=&\sum_{k=1}^nb_k(r)+c_n
\end{eqnarray}
The dipole equation is
\begin{equation}
  (1-\alpha_k(r))b_k(r)+\alpha_{k+1}(r)b_{k+1}(r)=0.
\label{e404}
\end{equation}
The factor $\displaystyle{\frac{e^{ar}}{r}}$ is monotonically increasing 
and $\cos x$ is periodic function plus or minus repeatedly, 
so the dipole equation will be followings replacing periodic summations 
to periodic integrals,
\begin{equation}
  \int_{(M+k)\pi+\beta_k(r)}^{(M+k+1)\pi}\frac{e^{ar}\cos r}{r}dr
  +\int_{(M+k+1)\pi}^{(M+k+1)\pi+\beta_{k+1}(r)}\frac{e^{ar}\cos r}{r}dr=0 
\end{equation}
associated with 
\begin{equation}
  \int_{(M+k)\pi+\beta_k(r)}^{(M+k+1)\pi+\beta_{k+1}(r)}
    \frac{e^{ar}\cos r}{r}dr=0.
\label{e406}
\end{equation}
\begin{equation}
  \frac
  {\displaystyle{\int_{(M+k)\pi+\beta_k(r)}^{(M+k+1)\pi}\frac{e^{ar}\cos r}{r}dr}}
  {\displaystyle{\int_{(M+k)\pi}^{(M+k+1)\pi}\frac{e^{ar}\cos r}{r}dr}}
  =1-\alpha_k(r).
\label{e407}
\end{equation}
\begin{equation}
  \frac
  {\displaystyle{\int_{(M+k+1)\pi}^{(M+k+1)\pi+\beta_{k+1}(r)}\frac{e^{ar}\cos r}{r}dr}}
  {\displaystyle{\int_{(M+k+1)\pi}^{(M+k+2)\pi}\frac{e^{ar}\cos r}{r}dr}}
  =\alpha_{k+1}(r).
\label{e408}
\end{equation}
where $\beta_k(r)$ is the internally dividing point in $k$-th sequence. 
Using Euler's formula\quad$\cos r+i\sin r=e^{ir}$,
\begin{eqnarray}
  \frac{e^{ar}\cos r}{r}
  &=&\Re(\frac{e^{ar}e^{ir}}{r})\nonumber\\
  &=&\Re(\frac{e^{cr}}{r})
\label{e409}
\end{eqnarray}
where $c=a+i$.
\begin{eqnarray}
  \int\frac{e^{cr}}{r}dr
  &=&\int\frac{1}{r}\sum_{k=0}^\infty\frac{cr^k}{k!}dr\nonumber\\
  &=&\log r+\frac{1}{c}\sum_{k=1}^\infty\frac{(cr)^{k+1}}{k!k}
\label{e410}
\end{eqnarray}
We express the real 
part of $\displaystyle{\int\frac{e^{cr}}{r}dr}$ 
by $E_\Re(r)$, 
then
\begin{eqnarray}
  E_\Re(r)
  &=&\Re\left(\int\frac{e^{cr}}{r}dr\right)\nonumber\\
  &=&\log r+\Re\left\{\frac{a-i}{a^2+1}
  \sum_{k=1}^\infty\frac{1}{k!k}(\sum_{m=0}^{k+1}{}_{k+1}C_{m}i^ma^{k+1-m})r^{k+1}\right\}
\label{e411}
\end{eqnarray}
We will get following equations from 
Eqs.(\ref{e406}) to (\ref{e408}), respectively
\begin{equation}
  E_\Re((M+k+1)\pi+\beta_{k+1}(r))=
  E_\Re((M+k)\pi+\beta_{k}(r))
\end{equation}
\begin{equation}
  \frac
  {E_\Re((M+k+1)\pi)-E_\Re((M+k)\pi+\beta_k(r))}
  {E_\Re((M+k+1)\pi)-E_\Re((M+k)\pi)}
  =1-\alpha_k(r)
\end{equation}
\begin{equation}
  \frac
  {E_\Re((M+k+1)\pi+\beta_{k+1}(r))-E_\Re((M+k+1)\pi)}
  {E_\Re((M+k+2)\pi)-E_\Re((M+k+1)\pi)}
  =\alpha_{k+1}(r)
\end{equation}

Solving $\beta_k(r)$ from these three equations, we can get $B_n(s,t)$ 
by integrating $\displaystyle{\frac{e^{ar}\cos r}{r}}$ 
until $n$ using obtained $\beta_k(r)$, and 
$B(s,t)$ will give us a regularized value by
\begin{equation}
  \beta(r)=\lim_{n:\beta_k(r)\rightarrow+\infty}\beta_n(r).
\end{equation}

Here we will show the concrete form of $\beta_n(r)$. The regularized value for
\begin{equation}
  E_n=\sum_{m=1}^n(-1)^m\frac{e^{am}}{m}=\sum_{m=1}^n\frac{(-e^a)^m}{m}
\end{equation}
will be got by the method of the dipole cancellation limit. 

The dipole equation will be
\begin{equation}
  (1-\alpha_k(r))\frac{e^{ak}}{k}-\alpha_{k+1}(r)\frac{e^{a(k+1)}}{k+1}=0,
\end{equation}
namely,
\begin{equation}
  e^a\frac{\alpha_{k+1}(r)}{k+1}+\frac{\alpha_k(r)}{k}=\frac{1}{k}
\label{e421}
\end{equation}
Then the dipole equation will be
\begin{equation}
  f(n)=\frac{1}{(-e^a)^{n-1}}f(1)
      -\frac{1}{(-e^a)^{n-1}}\sum_{k=1}^{n-1}\frac{(-e^a)^{k-1}}{k},
\end{equation}
where $f(k)=\displaystyle{\frac{\alpha_k(r)}{k}}$. So the solution $\alpha_n(r)$ is given by
\begin{eqnarray}
  \alpha_n(r)
  &=&-\frac{n}{(-e^a)^{n-1}}
     \left(\sum_{k=1}^{n-1}\frac{(-e^a)^{k-1}}{k}-\alpha_1\right)\nonumber\\
  &=&-\frac{n}{(-e^a)^{n-1}}\left(E_{n-1}\frac{1}{(-e^a)}-\alpha_1\right)\\
  &=&-\frac{n}{(-e^a)^n}\left(E_{n-1}+e^a\alpha_1\right)\nonumber,
\end{eqnarray}
wherre
\begin{eqnarray}
  E&=&\lim_{n\rightarrow\infty}E_{n-1}\nonumber\\
   &=&\lim_{n\rightarrow\infty}\sum_{m=1}^{n-1}\frac{(-e^a)^m}{m}\\
   &=&-\log(1+e^a)\nonumber
\end{eqnarray}

In this way, the $\alpha_1(r)$ will be
\begin{equation}
  \alpha_1(r)=\frac{-1}{e^a}(-\log(1+e^a))=\frac{1}{e^a}\log(1+e^a)
\end{equation}
as the dipole limit value 
and the regularized function is differentiable for $s$ and $t$. 

The identity theorem guarantees that the method of the dipole cancellation limit here 
is the unique analytic continuation for the standard form of the Riemann zeta function.
Moreover the function equality
\begin{equation}
  \zeta(z)=\pi^{z-\frac{1}{2}}\frac{\Gamma(\displaystyle{\frac{1-z}{2}})}
    {\Gamma(\displaystyle{\frac{z}{2}})}\zeta(1-z)
\end{equation}
gives us that the finiteness of the standard form for $0<s<1$. 
After all we can conclude that nontrivial zeros of the Riemann zeta function exist 
only on $s=1$, namely, on $\displaystyle{\Re z=\frac{1}{2}}$ for $0<\Re z<1$.

\section*{V. PRIME NUMBER FORMULA}
\hspace{\parindent}
As we mention details in the next section, the sufficient condition 
for the proof of the Riemann hypothesis is that 
the Euler product representation exists for the zeta function 
and the prime number theorem is satisfied. 
When this sufficient condition is satisfied, 
the prime number formula will be given in the category of elementary functions.
This prime number formula differs from such as 
the Wilson theorem\cite{Ribenboim} 
$$
  (p-1)!+1\equiv 0\;\;({\rm mod}\;p)\quad
  \Leftrightarrow\quad p{\rm \;is\;prime\;number}, 
$$
and the $n$-th prime number $p_n$ is extracted from the factor 
of the zeta function for the given $p_1, p_2, \cdots, p_{n-1}$ 
as follows:\cite{Fujimoto}
\begin{equation}
  p_n=\left[\left\{\log \zeta(a_n)
     +\sum_{r=1}^{n-1}\log\left(1-\frac{1}{{p_r}^{a_n}}\right)\right\}^{-1/a_n}\right]+1,
\label{e501}
\end{equation}
\noindent
where $[\;\;]$ is a Gauss symbol, 
and $a_n$ is a real number satisfying $a_n\ge p_n$.
When $n=1$, $\displaystyle{\sum_{r=1}^{n-1}}$ is equal to $0$．

  Taking the logarithmic function of each side of the Euler product representation of 
the zeta function $\zeta(s)=\displaystyle{\prod_{r=1}^\infty(1-{p_r}^{-s})^{-1}}$ and using 
$\log(1-x)^{-1}=\displaystyle{\sum_{k=1}^\infty\frac{x^k}{k}}\;\;\;(|x|<1)$, 
we get
\begin{equation}
\log\zeta(s)+\sum_{r=1}^{n-1}\log(1-{p_r}^{-s})=\sum_{k=1}^\infty\frac{1}{k}\sum_{r=n}^\infty {p_r}^{-ks}
\label{e502}
\end{equation}
So in order to prove Eq.(\ref{e501}), we can only show that 
the $(-1/a_n)$-th power of Eq.(\ref{e502}) exists in the interval 
$[\,p_n-1,p_n)$ for $s=a_n$. Namely we show 
\begin{equation}
\frac{1}{(p_n-1)^{a_n}}\ge\sum_{k=1}^\infty\frac{1}{k}\sum_{r=n}^\infty {p_r}^{-ka_n}
>\frac{1}{{p_n}^{a_n}}.
\label{e503}
\end{equation}
About the inequality between the middle and the right side, 
they are clear for
$$
  \sum_{k=1}^\infty\frac{1}{k}\sum_{r=n}^\infty {p_r}^{-ka_n}
  >\sum_{r=n}^\infty{p_r}^{-a_n}
  >{p_n}^{-a_n}\;\;\;(k=1, r=n).
$$
Next we show the inequality between the left and middle side.
\begin{eqnarray}
&&\sum_{k=1}^\infty\frac{1}{k}\sum_{r=n}^\infty {p_r}^{-ka_n}\nonumber\\
&<&\sum_{r=n}^\infty {p_r}^{-a_n}+\sum_{k=2}^\infty\frac{1}{2}\sum_{r=n}^\infty {p_r}^{-ka_n}\nonumber\\
&=&\sum_{r=n}^\infty {p_r}^{-a_n}+\sum_{r=n}^\infty\frac{1}{2{p_r}^{2a_n}(1-{p_r}^{-a_n})}\\
&<&\sum_{r=n}^\infty {p_r}^{-a_n}+\sum_{r=n}^\infty{p_r}^{-2a_n}<\sum_{m=p_n}^\infty m^{-a_n}\nonumber\\
&<&\int_{p_n-1}^\infty t^{-a_n}dt=\frac{p_n-1}{(a_n-1)(p_n-1)^{a_n}}\nonumber
\label{e504}
\end{eqnarray}
Thus accordingly if $a_n\ge p_n$ then $(p_n-1)/(a_n-1)\le 1$，
the right side of Eq.(80)$\le 1/(p_n-1)^{a_n}$. 
Thus the proof is completed.

  We can easily confirm the case of $n=1, a_1=2$，since $\zeta(a_1)=\zeta(2)=\pi^2/6$, we get
$$
  p_1=[\{\log\zeta(2)\}^{-1/2}]+1=2.
$$

Generally when $a_n$ is even，$\zeta(a_n)$ is expressed by the Bernoulli number. 
And the way to calculate the Bernoulli number is given by 
\begin{equation}
B_n=(-1)^n\sum_{m=0}^n\frac{1}{m+1}\sum_{t=0}^m(-1)^l{}_mC_n\,l^n
\end{equation}
for double summations or 
\begin{equation}
B_n=\frac{1}{2(2^{2n}-1)}\left(1+\left[\frac{2(2^{2n}-1)(2n)!}{2^{2n-1}\pi^{2n}}
\sum_{m=1}^{3n}\frac{1}{m^{2n}}\right]\right)
\end{equation}
for a single summation.\cite{Chowla} 

For instance, put $a_n=n(n+1)\;\;(n\ge 1)$ or 
$n=2[2^{-1}n\log(n\log n)]\;\;(n\ge 6)$\cite{Massias} 
in Eq.(\ref{e501}), and we can get $p_1, p_2, \cdots$, sequentially 
for all prime numbers.
By the proof above it is also clear that
\begin{equation}
  p_n=\lim_{s\rightarrow\infty}\left\{\log\zeta(s)
      +\sum_{r=1}^{n-1}\log\left(1-\frac{1}{{p_r}^s}\right)\right\}^{-1/s}
\end{equation}

The way to get a large prime number is discussed by the Hurwitz numbers. 
First we define the Hurwitz pi by 
\begin{equation}
 \varpi=2\int_0^1\frac{dx}{\sqrt{1-x^4}},
\label{e508}
\end{equation}
and Weierstrass's elliptic function $\wp(z)$ is defined by
\begin{equation}
  \wp(z)=\frac{1}{z^2}+\sum_{m^2+n^2\ne 0}
  \left\{\frac{1}{(z-\Omega_{m,n})^2}-\frac{1}{{\Omega_{m,n}}^2}\right\},
\end{equation}
where $\Omega_{m,n}=2m\omega_1+2n\omega_3$ with periods $2\omega_1$ and $2\omega_3$.
The Laurent expansion of $\wp(z)$ at the origin is given by
\begin{equation}
  \wp(z)=\frac{1}{z^2}+\sum_{n=0}^\infty\frac{2^nH_n}{n}\frac{z^{n-2}}{(n-2)!},
\end{equation}
where $H_n$ is the Hurwitz number which is not equal to zero only when $n$ is multiples of 4.The value of $H_4$ is equal to $\displaystyle{\frac{1}{10}}$ and the other can be got by 
the following recurrence formula: 
\begin{equation}
(2n-3)(4n-1)(4n+1)H_{4n}=3\sum_{i=0}^{n-1}(4i-1)(4n-4i-1){}_{4n}C_{4i}H_{4i}H_{4(n-i)}.
\end{equation}
We can get the $n$-th prime number $p_n$ in the form of $4m+1$ ($m$: integer) 
which is easily taken for large $m$, because the zeta function 
is expressed by the Hurwitz number,
\begin{equation}
  \zeta(a_n)=\frac{(2\varpi)^{4n}}{(4a_n)!}H_{4n},
\end{equation}
where $a_n$ is multiples of 4.
In this way we proceed the relation between the generalized Bernoulli number 
and the generalized Hurwitz pi, which is given by the higher power of $x$ 
of the integrand in Eq.(\ref{e508}), for instance, 
$\displaystyle{2\int_0^1\frac{dx}{\sqrt{1-x^{2^n}}}}$ 
in Hurwitz way to get large prime numbers.

\section*{VI. CONCLUSIONS AND REMARKS}
\hspace{\parindent}
Considering what are stated in above sections, 
Riemann hypothesis will be realized when
\begin{enumerate}
  \item An Euler product representation exists such as
    \begin{equation}
      \zeta(z)=h_1(z)\prod_p(1-\frac{1}{p^z})^{-1}.
    \label{e601}
    \end{equation}
  \item The function of $z$, namely, an infinite summation of prime numbers $c(z)$,
    \begin{equation}
      c(z)=\sum_{k=1}^\infty\frac{1}{{p_k}^z},
    \label{e602}
    \end{equation}
    converges for $\Re z>1$.
  \item The function equality is satisfied such as
    \begin{equation}
      \zeta(z)=h_2(z)\zeta(1-z).
    \label{e603}
    \end{equation}
  \item For zeros existing on $\Re z=\displaystyle{\frac{1}{2}}$,
    \begin{equation}
      c(1)=\sum_{k=1}^\infty\frac{1}{p_k}=+\infty
    \label{e604}
    \end{equation}
must be satisfied.
\end{enumerate}

The first condition, {\it i.e.}, the Euler product representation played the role 
as stated above, and it will be extended to the condition
\begin{equation}
  \zeta(z+iu)=h_3(z)\prod_p\left(1-\frac{\chi}{p^{az}}\right)^{-1},
\label{e605}
\end{equation}
to get the proofs of the Riemann hypotheses using the corresponding 
$B(s,t)$ in \S3 through \S5, 
where $a(>0)$, $u$ and $\chi(\ne 0)$ are real constants, and $h_3(z)$ is a real function.
The second and forth condition are easily shown to be concluded when 
the generalized prime number theorem
\begin{equation}
  \pi(n)=O\left(\frac{n}{\log n}\right).
\label{e606}
\end{equation}
is satisfied. 
Therefore the zeta functions, which are satisfied with Eqs.(\ref{e605}),(\ref{e606}) 
and the third condition, make their Riemann hypotheses to be proved. 
Especially the Dirichlet $L$-function is satisfied with these three conditions, 
zeros of the $L$-function exist only on the straight line of 
$\displaystyle{\Re z=\frac{1}{2}}$ except zeros of $h_3(z)=0$.

\vskip 10mm
\noindent
{\bf ACKNOWLEDGEMENTS}

  One of the authors(M.F.) would like to express his gratitude for the hospitality at 
Tezukayama University where many invaluable discussions were made on this work.

\vskip 5mm
\newpage
\renewcommand{\theequation}{\Alph{section}\arabic{equation}}
\setcounter{section}{1}
\setcounter{equation}{0}
\section*{APPENDIX \Alph{section}}

\hspace{\parindent}
Here we apply the regularization method stated in \S 2 to the Riemann zeta function 
$\zeta(z)=\displaystyle{\lim_{n\rightarrow\infty}\zeta_n(z)}
=\displaystyle{\lim_{n\rightarrow\infty}\sum_{k=1}^n\frac{1}{k^z}}$.
The dipole equation Eq.(\ref{e205}) is
\begin{equation}
  \{1-\alpha_k(z)\}\frac{1}{k^z}+\alpha_{k+1}(z)\frac{1}{(k+1)^z}=0.
\label{a01}
\end{equation}
So we solve the equation
\begin{equation}
 \frac{\alpha_k(z)}{k^z}-\frac{\alpha_{k+1}(z)}{(k+1)^z}=\frac{1}{k^z},
\label{a02}
\end{equation}
and we get the solution given by
\begin{equation}
  \alpha_1(z)=\lim_{n\rightarrow\infty}\left\{\sum_{k=1}^{n-1}\frac{1}{k^z}
  +\frac{\alpha_n(z)}{n^z}\right\}.
\label{a03}
\end{equation}

The expression $\displaystyle{\alpha_k(z)=\frac{k}{1-z}}$ is satisfied 
with Eq.(\ref{a02}) in the limit of $k\rightarrow+\infty$.
Thus the regularized zeta function is
\begin{equation}
 \zeta(z)=\lim_{n\rightarrow\infty}\left\{\zeta_n(z)-\frac{n^{1-z}}{1-z}\right\},
\end{equation}
which is differentiable except $z=1$ and analytically continued to the region 
for $\Re z>0$. 

This result agrees with the analytic continuation 
by the usual Euler-Maclaurin expansion. 
Here we show that this is consistent with Eq.(\ref{e102}) 
again by way of the method of the dipole cancellation limit.
We put $\displaystyle{\xi(z)=\lim_{n\rightarrow\infty}\xi_n(z)}$, where
\begin{equation}
 \xi_n(z)=\sum_{k=1}^n\frac{(-1)^{n-1}}{n^z}.
\end{equation}
By applying the method of the dipole cancellation limit to $\xi_n(z)$, 
we get the dipole equation
\begin{equation}
  \{1-\alpha_k(z)\}\frac{1}{k^z}+\alpha_{k+1}(z)\}\frac{(-1)}{k^{z+1}}=0.
\end{equation}
The solution of this equation is as follows:
\begin{equation}
  \alpha_1(z)=\sum_{k=1}^{n-1}\frac{(-1)^{k-1}}{k^z}+(-1)^{n-1}f_n(z),
\end{equation}
where $\displaystyle{f_k(z)=\frac{\alpha_k(z)}{k^z}}$.

For the odd number $2n+1$, $\xi_{2n+1}(z)$ is expressed by $\xi_{2n}(z)$
\begin{eqnarray}
  \xi_{2n+1}(z)&=&\xi_{2n}(z)-\frac{1}{(2n+1)^z}\nonumber\\
  &=&\xi_{2n}(z)+O\left(\frac{1}{(2n+1)^z}\right),
\end{eqnarray}
where last term vanishes in the limit $n\rightarrow\infty$. So it is 
sufficient to demonstrate for the even case of $\xi_{2n}(z)$.
For the even number $2n$, $\xi_{2n}(z)$ can be shown the relation to 
the Riemann zeta function, 
\begin{eqnarray}
  \xi_{2n}(z)&=&\zeta_{2n}(z)-\frac{2}{2^z}\zeta_n(z)\nonumber\\
  &=&\zeta_n(z)+\sum_{k=n+1}^{2n}\frac{1}{k^z}-\frac{2}{2^z}\zeta_n(z)\\
  &=&(1-2^{1-z})\zeta_n(z)+\sum_{k=n+1}^{2n}\frac{1}{k^z}.\nonumber
\end{eqnarray}
The last summation term can be evaluated in the limit of $n\rightarrow\infty$ 
as follows,
\begin{eqnarray}
  \lim_{n\rightarrow\infty}\sum_{k=n+1}^{2n}\frac{1}{k^z}
  &=&\lim_{n\rightarrow\infty}\sum_{k=1}^n
    \frac{1}{(1+\frac{k}{n})^z\,n^{z-1}}\frac{1}{n}\nonumber\\
  &=&\lim_{n\rightarrow\infty}\sum_{k=1}^n
    n^{1-z}\left(\frac{1}{(1+\frac{k}{n})^z}\,\frac{1}{n}\right)\nonumber\\
  &=&\lim_{n\rightarrow\infty}n^{1-z}\int_1^2\frac{1}{t^z}dt\\
  &=&\lim_{n\rightarrow\infty}n^{1-z}\left[\frac{t^{1-z}}{1-z}\right]_1^2\nonumber\\
  &=&\lim_{n\rightarrow\infty}n^{1-z}
    \left(\frac{2^{1-z}-1}{1-z}\right).\nonumber
\end{eqnarray}

After all the limit of $\xi_{2n}(z)$ for $n\rightarrow\infty$ 
is expressed by the regularized zeta function
\begin{eqnarray}
  \lim_{n\rightarrow\infty}\xi_{2n}(z)
  &=&\lim_{n\rightarrow\infty}(1-2^{1-z})
    \left\{\zeta_n(z)-\frac{n^{1-z}}{1-z}\right\}\nonumber\\
  &=&(1-2^{1-z})\lim_{n\rightarrow\infty}
    \left\{\zeta_n(z)-\frac{n^{1-z}}{1-z}\right\}\\
  &=&(1-2^{1-z})\zeta(z),\nonumber
\end{eqnarray}
namely, Eq.(\ref{e102}) is obtained.

\vskip 5mm
\addtocounter{section}{1}
\setcounter{equation}{0}
\section*{APPENDIX \Alph{section}}

\hspace{\parindent}
\indent
In the case of $\displaystyle{\lim_{n\rightarrow\infty}A_n(x_i)}=\infty$, 
Eq.(\ref{e313}) is manifestly divergent in each side. 
So we show the equlity only in the case that 
$\displaystyle{\lim_{n\rightarrow\infty}A_n(x_i)}$ is finite.
In this case there exists finite number $M$ such that
$\displaystyle{\lim_{n\rightarrow\infty}A_n(x_i)<M}$,
we express $A_n(x_i)$ by $A_{n,M}(x_i)$ explicitly and
$$
g_{n,M}(x_i)=\left(1+\frac{A_{n,M}(x_i)}{n}\right)^n,\;\;\;
g_M(x_i)=\lim_{n\rightarrow\infty}g_{n,M}(x_i),
$$
\noindent
then taking logarithm function of this $g_M(x_i)$ and setting $n=h^{-1}$, we get
\begin{eqnarray}
\log g_{h^{-1},M}(x_i)
&=&\frac{\log\{1+hA_{h^{-1},M}(x_i)\}}{h}\nonumber\\
&=&\frac{1}{h}\sum_{k=1}^\infty\frac{(-1)^{k-1}}{k!}\{hA_{h^{-1},M}(x_i)\}^k\\
&=&A_{h^{-1},M}(x_i)+\sum_{k=2}^\infty\frac{(-1)^{k-1}}{k!}h^{k-1}A_{h^{-1},M}(x_i)^k\nonumber
\end{eqnarray}  
The Taylor expansion for the logarithmic function can be done, because $h$ can be 
taken small enough for $|A_{h^{-1},M}(x_i)|<M$ satisfying $|hA_{h^{-1},M}(x_i)|<1$.
Taking the limit of $n\rightarrow\infty$
\begin{eqnarray}
\log g_M(x_i)&=&\log(\lim_{h\rightarrow 0}g_{h^{-1},M}(x_i))
=\lim_{h\rightarrow 0}\log g_{h^{-1},M}(x_i)\nonumber\\
&=&\lim_{h\rightarrow 0}\left\{A_{h^{-1},M}(x_i)
+\sum_{k=2}^\infty\frac{(-1)^{k-1}}{k!}h^{k-1}A_{h^{-1},M}(x_i)^k\right\}\nonumber\\
&=&\lim_{h\rightarrow 0}A_{h^{-1},M}(x_i)
+\sum_{k=2}^\infty\frac{(-1)^{k-1}}{k!}
\left\{\lim_{h\rightarrow 0}h^{k-1}A_{h^{-1},M}(x_i)^k\right\}\\
&=&\lim_{h\rightarrow 0}A_{h^{-1},M}(x_i)+0=A_M(x_i).\nonumber
\end{eqnarray}
Thus $g_M(x_i)=e^{A_M(x_i)}$ is shown.

\vskip 5mm

\end{document}